\documentclass{elsart}

\usepackage[numbers]{natbib}

\usepackage{graphicx}

\usepackage{amssymb}
\usepackage{amsfonts}

\graphicspath{%
    {converted_graphics/}
    {D:/thermal_conductivity/}
    {D:/thermal_conductivity/figs/}
}

\newcommand{\spider}{{\sc Spider} }

\begin{document}
\bibliographystyle{plainnat}

\begin{frontmatter}



\title{Thermal Conductivity of Thermally-Isolating Polymeric and Composite
Structural Support Materials Between 0.3 and 4~K}


\author[Caltech]{M.C. Runyan},
\author[JPL,Caltech]{W.C. Jones}

\address[Caltech]{Department of Physics, California Institute of Technology, MC59-33, 1201 E. California Blvd., Pasadena, CA 91125}
\address[JPL]{Observational Cosmology Group, Jet Propulsion Laboratory,
4800 Oak Grove Drive, Pasadena, CA 91109 }

\begin{abstract}
We present measurements of the low-temperature thermal conductivity of a number of polymeric and composite materials from 0.3 to 4~K.  The materials measured are Vespel SP-1, Vespel SP-22, unfilled PEEK, 30\% carbon fiber-filled PEEK, 30\% glass-filled PEEK, carbon fiber Graphlite composite rod, Torlon 4301, G-10/FR-4 fiberglass, pultruded fiberglass composite, Macor ceramic, and graphite rod.  These materials have moderate to high elastic moduli making them useful for thermally-isolating structural supports.     

\end{abstract}

\begin{keyword}
Thermal conductivity \sep Polymers \sep Composites \sep Structural materials
\end{keyword}

\end{frontmatter}

\section{Introduction}

Cryogenic instruments frequently require rigid mechanical support of significant mass while minimizing conductive heat flow.
In particular, optical systems require very stiff structural mounts.  To increase the stiffness of a structure the designer generally has the choice of using more material or choosing materials with larger elastic moduli.   
However, materials with higher elastic moduli tend to have higher thermal conductivity.  This situation motivates the use of the ratio of elastic modulus to thermal conductivity as the figure of merit for thermally-isolating structural materials. 

This work was motivated by the need to support the cryogenic microwave receiver of the \spider balloon experiment\citep{montroy2006,mactavish2008,crill2008}.  The \spider instrument insert has massive components at 4.2, 1.4, 0.35 and 0.25~K.  Each of these temperature stages are thermally and structurally referenced to an outer liquid helium cryostat.  The cryostat will tip in elevation from zenith to horizon and the cryogenic support structures must minimize the gravitational deflection of the focal plane.  
Historically, we have used polyimide supports such as Dupont Vespel SP because
of its low thermal conductivity \citep{locatelli1976,olson1993}.  Advanced
polymers like Vespel are generally quite expensive.  At the time of this
writing, 1/2$^{\prime\prime}$ Vespel rods cost approximately half the price of gold by weight.  
Because the \spider instrument will include six duplicate
focal plane structures, the cost and availability of the material required to
support them is a significant concern.

There are a number of materials properties that are potentially important when
designing thermally-isolating structural supports.  As mentioned previously, low thermal conductivity and high modulus are desirable attributes.  The elastic modulus of materials increases as they cool (for example, see \citet{flynn}).  This increase in stiffness is moderate in metals but can be significant for for some polymers.  In particular, the elastic modulus of Teflon increases by approximately a factor of 20 when cooled from room to cryogenic temperatures \citep{corruccini1957}.  In addition to stiffness, the strength of materials is important for structures that may experience large forces or shocks.

The stability of the dimensions of the structure upon cooling may also be important.  In this case, materials with small coefficients of thermal contraction, such as carbon fiber composites \citep{reed1997}, may be desirable. For larger structures the support members themselves may comprise a significant fraction of the mass, in which case the density may become important.  The enthalpy of the materials, particularly those with low thermal conductivity, may also be a significant concern since they will take longer to cool down.      
  
\section{Materials Tested}
We test two types of DuPont Vespel polyimides.  Vespel SP-1 is the base polyimide resin and SP-22 is filled with 40\% graphite by weight. Polyetheretherketone (PEEK) is a semi-crystalline thermoplastic and is available in both pure and filled forms.  We test three types of PEEK. The first is unfilled Ketron PEEK 1000 manufactured by Quadrant Plastics\footnote{http:{\slash}{\slash}www.quadrantepp.com}. The second is 30\% carbon fiber-filled PEEK 450CA30 manufactured by Drake Plastics\footnote{http:{\slash\slash}www.drakeplastics.com} (PEEK CA30).  The short carbon fibers are not oriented in this material, although the extrusion process may introduce some anisotropy to its properties.  We only measure the thermal conductivity along the extrusion direction of the rod. The third PEEK sampe is 30\% glass-filled Sustatec PEEK manufactured by Rochling Sustaplast\footnote{http:{\slash\slash}www.sustaplast.com} (PEEK GF30). This is also an extruded product and we measured the thermal conductivity only along the axis of extrusion.

We measure two composite materials made by Avia Sport
Composites\footnote{http:{\slash\slash}www.aviasport.net}.  The first is
pultruded carbon-fiber called Graphlite.  Graphlite contains 67\% carbon fiber
by volume and a Bis-F epoxy matrix.  The pultrusion process naturally orients the fiber bundle along the axis of the rod, but visual inspection of the material shows that the fibers are not uniformly oriented along this axis.  The second material is a pultruded fiberglass rod with Bis-F epoxy which we will refer to as Avia Fiberglass in this paper.  We only measure the thermal conductivity along the pultrusion axis of the rods.  Both of these composites are supplied by CST Composites\footnote{http:{\slash\slash}www.cstsales.com}.

We measure a sample of graphite rod from Poco Graphite Inc.\footnote{http:{\slash\slash}www.poco.com}  The type of graphite measured is the same as that tested in \citet{woodcraft2003} (industrial grade AXM-5Q).  This graphite has a particle size of 5 microns and an apparent density of 1.73 g/cc.  Macor is a machinable glass-ceramic manufactured by Corning\footnote{http:{\slash\slash}www.corning.com}.  We also measure a sample of Torlon 4301.  The Torlon base (4203) is a polyamide-imide (PAI) thermoplastic manufactured by Quadrant Plastics and the 4301 grade includes both 12\% graphite powder and 3\% PTFE.  

The final material tested is a commonly available G-10/FR-4 Garolite
fiberglass rod purchased from
McMaster-Carr\footnote{http:{\slash\slash}www.mcmaster.com}.
The manufacturer of this sample is unknown, but the distributor states that it meets MIL-I-24768 specification for G-10/FR-4.  Although a cryogenic grade of G-10 is available, our own experience and that of \citet{walker1981} suggests that for many cryogenic applications the readily available G-10/FR-4 is an acceptable material.  However, FR-4 may be unsuitable for ultra-high vacuum applications because of the presence of a halogen flame retardant.  These rods are fabricated by bonding layers of woven glass fabric into a sheet and then grinding them into rods along one of the fiber axes.  We confirmed the orientation of the fibers in our sample by dissolving away the binding epoxy.  We measured the thermal conductivity along the plane of the woven glass fabric.      

\begin{table}
\centering
\begin{tabular}{cccc}
\hline
Material & $\rho$ (g/cc) & {TM (GPa)} & CM (GPa)  \\
\hline\hline
Vespel SP-1 & 1.43 & $2.6^{\dagger}$, $2.2^{\ddagger}$ & 2.4  \\
Vespel SP-22 & 1.65 & $2.3^{\dagger}$, $3.3^{\ddagger}$ & 3.3 \\
PEEK & 1.31 & 4.3 & 3.5 \\
PEEK CA30${}^{**}$ & 1.41 & 7.7 & 4.9 \\
PEEK GF30 & 1.51 & 6.9 & - \\
Graphlite CF Rod & 1.55 & 134 & 131 \\
Avia Fiberglass Rod & 2.05 & 45 & -  \\
Torlon 4301 & 1.45 & 6.2 & 6.6  \\
G-10/FR-4 & 1.91 & - & -  \\
Macor & 2.52 & $67^*$ & $67^*$  \\
Poco Graphite AXM-5Q & 1.73 & $10.5^*$ & $10.5^*$  \\
\hline
\end{tabular} 
\caption{\scriptsize Room-temperature properties of materials tested. $\rho$ is the density, TM is the tensile modulus, CM the compression modulus.  For pultruded composite materials, the TM and CM are measured along the fiber direction.  All values are obtained from manufacturers' data sheets and websites unless otherwise stated. $\dagger$ - indicates value is estimated from manufacturer's stress-strain curve at 1\% strain.  $\ddagger$ - data from material properties table published by Oblicos-96, Ltd. (www.oblicos.com).  $*$ - indicates that only elastic modulus data available. $**$ - Mechanical data for PEEK CA30 is from the equivalent product Ketron PEEK CA30 manufactured by Quadrant Plastics.  The only reliable modulus data for G-10/FR-4 we could find was the flexure modulus: 18.6 GPa parallel and 16.5 GPa perpendicular to the fiberglass orientation. }
\label{matproptable}
\end{table}

\section{Experimental Procedure}
The thermal conductivity measurements are performed in an IR Labs HD-3(10)L liquid nitrogen/liquid helium cryostat.  Temperatures below 0.27~K are obtained with a ${}^3$He sorption fridge operating from the pumped helium bath of the cryostat.  The experiment is conducted within the 1.4~K radiative environment of the pumped helium cryostat.

\begin{figure}[t] 
  \centering
  \includegraphics[bb=36 101 756 528,width=5.67in,height=3.36in,keepaspectratio]{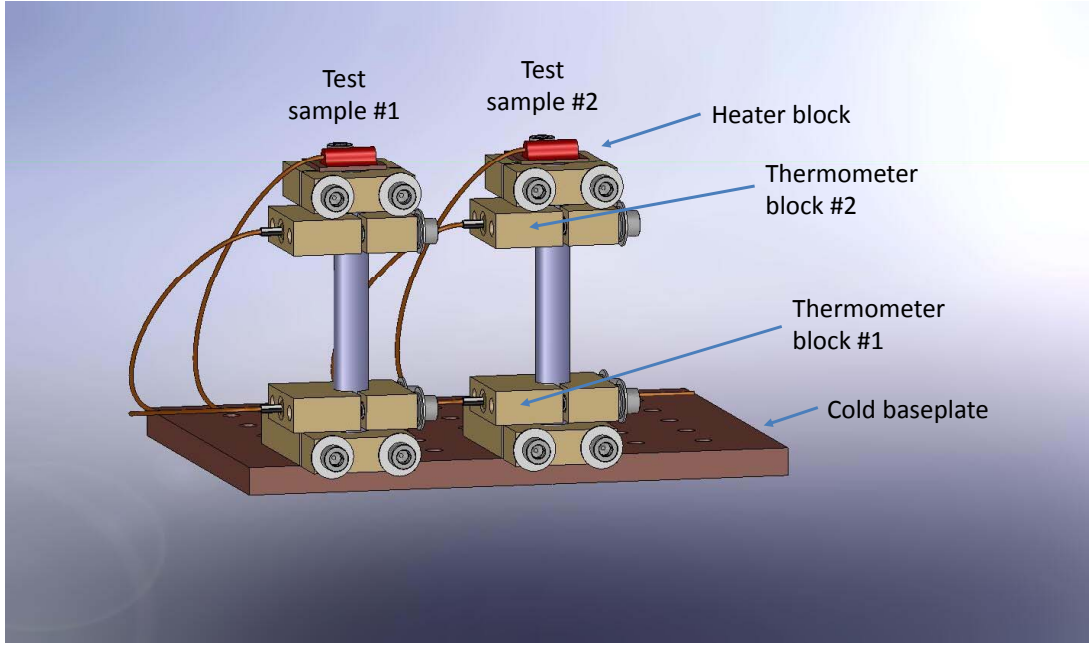}
  \caption{\scriptsize Schematic of the experimental setup.  There are two samples in the test dewar but they are measured individually.  The schematic shows the heater blocks, Cernox thermometer blocks and baseplate, which is attached to the sub-Kelvin ${}^3$He sorption fridge.}
  \label{fig:setup}
\end{figure}

The sample test stage (shown schematically in Figure \ref{fig:setup}) has
provision for two independent samples.  The C10100-alloy copper mounting
blocks are two piece clamps.  Because the coefficient of thermal contraction
of the test samples is generally higher than copper we use stainless
belleville spring washers to maintain clamping force.  A mounting block holds
the sample upright and thermally sinks it to the cold stage.  There are two
thermometer blocks along the length of the sample, each containing a
calibrated Cernox thermometer read out with a 4-wire resistance bridge.  At
the top of each sample is a 1 M$\Omega$ metal-film resistive heater.  In order
to track any temperature coefficient of the resistive heater, we monitor both
voltage and current across the heater to determine the heater power.  
Over the range of temperatures reported in this paper, the heater impedance is observed to vary by 7\% .
A thin layer of Apiezon-N thermal grease is applied to all interfaces.  The test samples were all turned down to 0.25$^{\prime\prime}$ in diameter with the exception of the Poco Graphite sample, which was 3/16$^{\prime\prime}$ diameter and shimmed in the mounting blocks. The overall length of each sample was $\sim$2$^{\prime\prime}$.  The two thermometer blocks are spaced approximately 1$^{\prime\prime}$ apart and their separation is measured with calipers.

An important element of the experimental setup is the separation of the
thermometers from the heaters and heatsinks.  When heat flows across a thermal
interface a temperature gradient is formed.  This thermal boundary resistance
can lead to a significant temperature offset between the thermometer block and
the sample under test if heat flows through the clamp to the material.  If the
thermometers are separated from the heaters and heatsinks, then once the test
sample has come to thermal equilibrium there is very little heat flow from the thermometer block into the material (of order several pico--Watts). 
We can therefore be confident that the thermometer is reading the temperature
of the material at that point along its length.  Failure to effectively
isolate the thermal boundary impedance from the intrinsic thermal conductivity
of the sample will result in a biased estimate of the thermal conductivity of the sample.    

\begin{figure}[tbp] 
  \centering
  \includegraphics[bb=88 84 544 522,width=5in,height=4.8in,keepaspectratio]{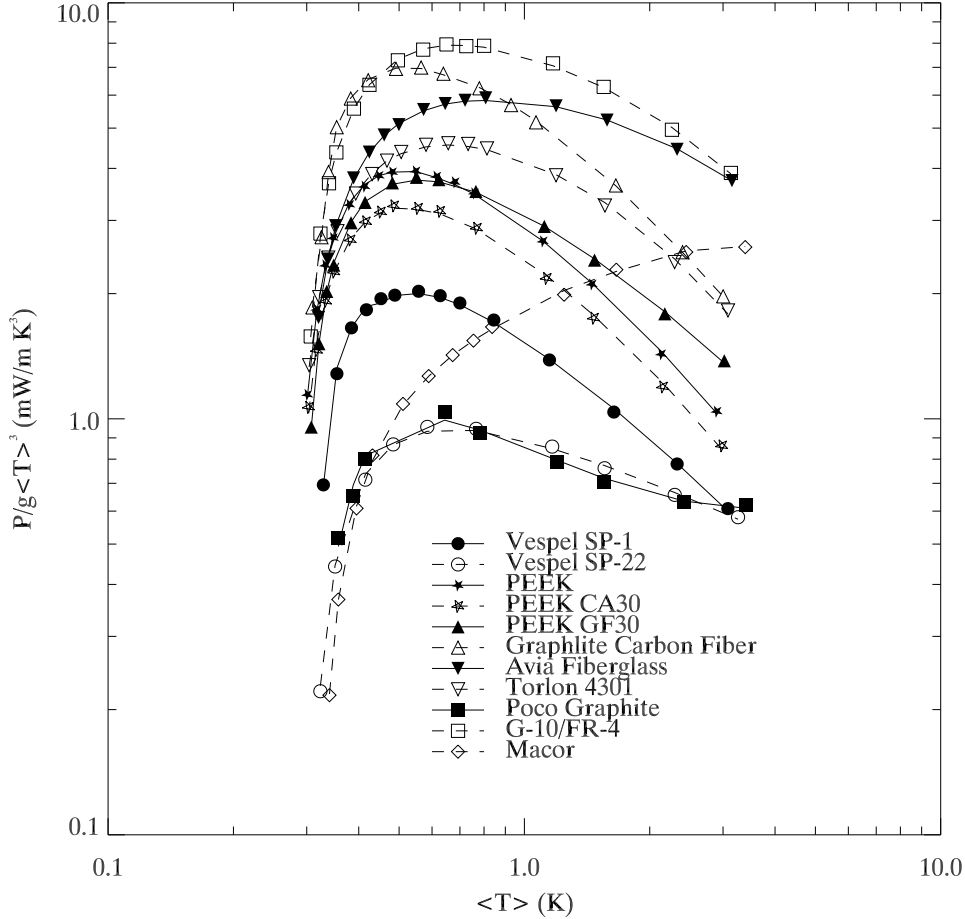}
  \caption{\scriptsize The data points are the applied heater power, $P$, divided by the geometric factor, $g=A/L$, for each material. We have plotted these versus the weighted temperature, $\langle T\rangle = \int_{T_L^i}^{T_H^i} T~k(T)~dT / \int_{T_L^i}^{T_H^i} k(T)~dT$, for each data point, $i$.  $k(T)$ is the best--fit thermal conductivity function for each material as discussed below.  We have divided the heater power by $\langle T\rangle^3$ to space the curves more clearly in the figure.  The lines show the heat flow for the best--fit thermal conductivity between each pair of temperature points, $\{T_L^i,T_H^i\}$.  The agreement between the data points and lines illustrates the quality of the the fit between the data and thermal conductivity model. }
  \label{fig:rawdata}
\end{figure}

We use the steady--state heat method and apply a measured amount of heater power to the top of each sample, wait until the system comes to thermal equilibrium, and measure the equilibrium temperatures at two points along the length of the sample.  The wiring used to read out thermometers and bias heaters is 0.0045$^{\prime\prime}$ diameter manganin. The thermal conductivity of the wiring is estimated to be no more than 2\% of the conductivity of the least conductive material.  The measurements are shown in Figure \ref{fig:rawdata}.

\section{Results and discussion}
The data are sparsely sampled between 0.27 and 4.2~K and the lower temperature
changes with applied power because of the thermal boundary impedance with the
heat sink. These factors prevent us from taking the derivative with respect to
the upper temperature and employing the Leibniz rule for differentiating
integrals.  
We instead fit the integral of a thermal conductivity model to the measured equilibrium temperatures and applied thermal power. 

The conductive heat flow through the test sample for the $i^{th}$ data point is
$$
P^i/g = \int_{T_L^i}^{T_H^i} k(T)~dT~,
$$
where $P^i$ is the heat flow, $g=A/L$ is the geometric factor, $k(T)$ is the thermal conductivity, $T_L^i$ is the lower temperature and $T_H^i$ the upper temperature.  We fit our data to a thermal conductivity model of the form
$$
k(T) = \alpha T^{\beta_{eff}(T)} = \alpha T^{(\beta+\gamma T^n)}~.
$$

The thermal conductivity of many of our samples begins to plateau at temperatures below $T\sim4$~K.  Using the functional form  
 $\beta_{eff}(T) = \beta + \gamma T^n$ allows a convenient parameterization of this rolling index.  The phenomenon of the thermal conductivity reaching a plateau is common in polymers \citep{freeman1986,duval1995,barucci2005}. It is important to note that there are degeneracies between the parameters of our thermal conductivity model -- the individual parameters are not physically motivated.  However, the $\beta_{eff}(T)$ should encode information about the physical processes involved at each temperature.

\begin{figure}[tbp] 
  \centering
  \includegraphics[bb=84 79 544 715,width=5in,height=6.92in,keepaspectratio]{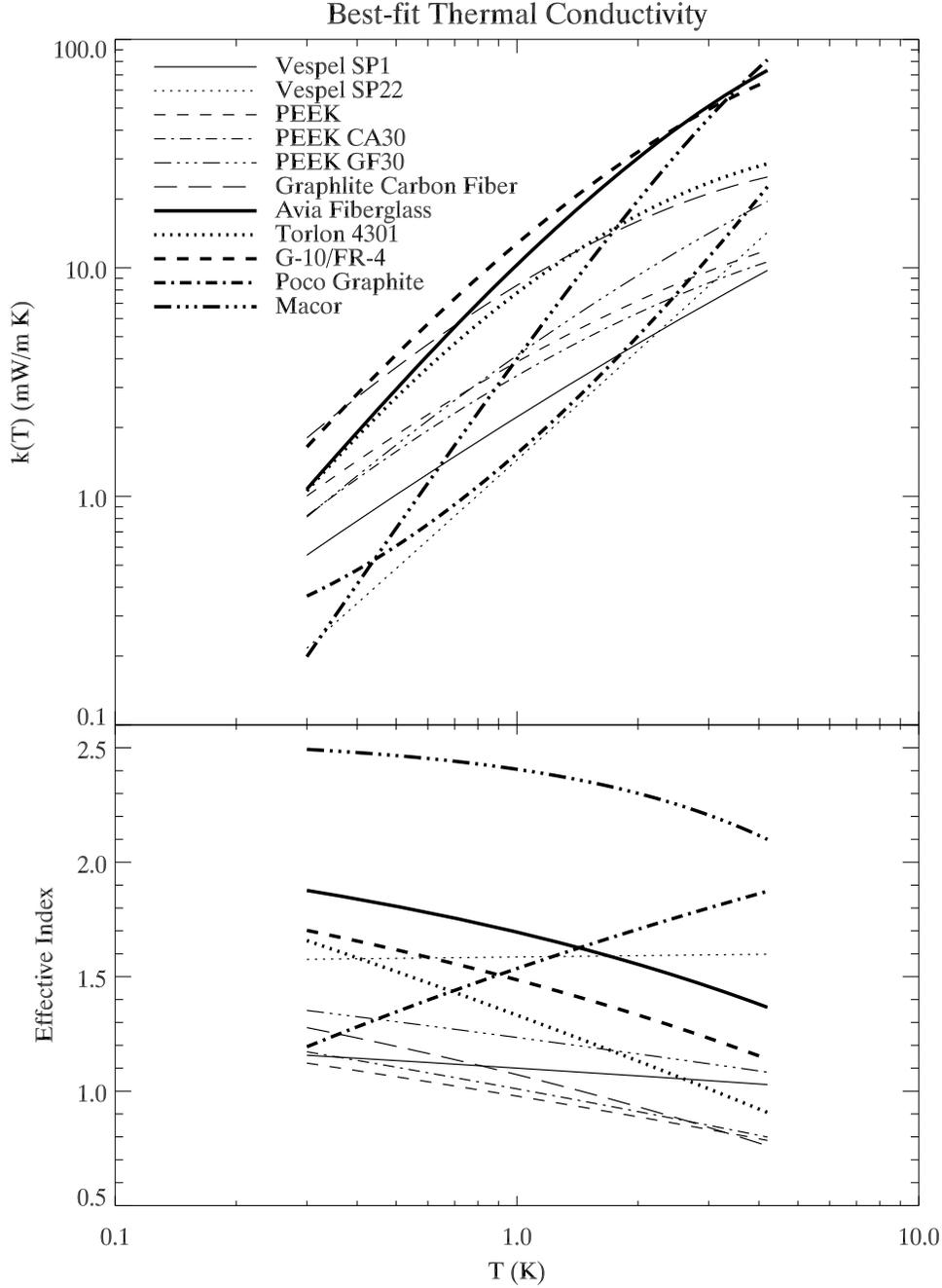}
  \caption{\scriptsize The top plot shows the best--fit thermal conductivity for the materials tested.  The bottom plot shows the effective index, $\beta_{eff}(T) = \beta + \gamma T^n$, for each material.}
  \label{fig:kt}
\end{figure}

For each material we use the set of electrical powers and temperatures and fit the set of data
$$
P_{elect}^i = g\int_{T_L^i}^{T_H^i} k(T)~dT + P_{offset}~,
$$
where $i$ represents the particular equilibrium point and $P_{offset}$ is a small constant--power offset.  $P_{offset}$ could include thermometer calibration error, wiring parasitics, and radiative coupling and is typically measured to be between 5 and 20~nW.  The results of the fits can be found in Table \ref{tbl:ktfits} and are valid for temperatures within the measured range of 0.3 to 4.2~K.  These functions are plotted in Figure \ref{fig:kt} as well as the effective index, $\beta_{eff}(T) = \beta + \gamma T^n$.  Note that because the variables used in this fit are not uncorrelated different combinations of $\alpha$, $\beta$, $\gamma$, and $n$ may result in $k(T)$ that are identical within the accuracy of the data.

\begin{table}
\centering
\begin{tabular}{ccccc}
\hline
Material & $\alpha$ (mW/m~K) & $\beta$ & $\gamma$ (1/K${}^n$) & $n$  \\
\hline\hline
Vespel SP--1             & 2.23 & 1.92 & -0.819 & 0.0589  \\
Vespel SP--22            & 1.44 & 2.11 & -0.521 & -0.0163 \\
PEEK                    & 3.88 & 2.41 & -1.43 & 0.0884 \\
PEEK CA30               & 3.37 & 3.20 & -2.19 & 0.0640 \\
PEEK GF30               & 4.14 & 3.07 & -1.84 & 0.0553 \\
Graphlite CF Rod        & 8.39 & 2.12 & -1.05 & 0.181 \\
Avia Fiberglass Rod     & 10.3 & 2.28 & -0.585 & 0.310 \\
Torlon 4301             & 7.77 & 5.46 & -4.13 & 0.0682 \\
G--10/FR--4               & 12.8 & 2.41 & -0.921 & 0.222 \\
Macor                   & 4.00 & 2.55 & -0.140 & 0.809 \\
Poco Graphite AXM--5Q    & 1.54 & 3.36 & -1.83 & -0.142 \\
\hline
\end{tabular} 
\caption{\scriptsize Best fit values for the thermal conductivity of materials in the temperature range of 0.3 to 4.2~K.  The functional form of the fit is $k(T) = \alpha T^{(\beta+\gamma T^n)}$.  As discussed in the text, the thermal conductivity of bulk fiber-containing composites is measured along the axis of the fibers.}
\label{tbl:ktfits}
\end{table}

\begin{table}
\centering
\begin{tabular}{cccccc}
\hline
Material & {\scriptsize $k(T=0.3$~K$)$} & {\scriptsize $k(T=1.4$~K$)$} & {\scriptsize $k(T=4.2$~K$)$} & {\scriptsize $\int_{0.3K}^{1.4K} k(T)~dT$} & {\scriptsize $\int_{1.4K}^{4.2K} k(T)~dT$} \\
  & {\scriptsize(mW/m~K)} & {\scriptsize(mW/m~K)} & {\scriptsize(mW/m~K)} & {\scriptsize(mW/m)} & {\scriptsize(mW/m)} \\
\hline\hline
\scriptsize Vespel SP--1             &  0.553 & 3.21 & 9.74 & 2.05 & 18.3 \\
\scriptsize Vespel SP--22            &  0.217 & 2.46 & 14.3 & 1.31 & 21.7 \\
\scriptsize PEEK                    &  1.00  & 5.31 & 12.0 & 3.57 & 25.3 \\
\scriptsize PEEK CA30               &  0.823 & 4.66 & 10.6 & 3.09 & 22.4 \\
\scriptsize PEEK GF30               &  0.812 & 6.20 & 19.6 & 3.76 & 36.7 \\
\scriptsize Graphlite CF Rod        &  1.80  & 11.8 & 25.0 & 7.62 & 55.0 \\
\scriptsize Avia Fiberglass Rod     &  1.08  & 17.8 & 73.2 & 9.17 & 130 \\
\scriptsize Torlon 4301             &  1.06  & 11.8 & 28.5 & 6.88 & 60.7 \\
\scriptsize G--10/FR--4               &  1.64  & 20.6 & 65.6 & 11.4 & 126 \\
\scriptsize Macor                   &  0.199 & 8.85 & 81.4 & 3.64 & 116 \\
\scriptsize Poco Graphite AXM--5Q    &  0.366 & 2.66 & 22.7 & 1.44 & 29.9 \\
\hline
\end{tabular} 
\caption{\scriptsize Best--fit values of the thermal conductivity of the materials tested at common cryogenic temperatures.  Also included are the integrals of the best--fit thermal conductivity functions between these temperatures.}
\label{tbl:ktvals}
\end{table}

We assign an uncertainty of 10\% to a value of thermal conductivity generated
using the parameters presented in Table \ref{tbl:ktfits} throughout the range
0.3 to 4~K.  The statistical scatter of the residuals to the fits is generally
of order a few percent indicating that the model is a good fit to the data.
We have not corrected for the change in $g=A/L$ due to the contraction of the
materials as they cool from room temperature and our measurements of $g$ were
of order 1\%.  The uncertainty on the heater power applied on the top of the
sample is estimated at 4\%.  We estimate the parasitic conductivity from the
wiring to be less than 2\% for the least conductive materials.  The
calibration of the thermometers is accurate to better than 1\% with respect to the calibration from Lakeshore Cryotronics.  We measured a PEEK CA30 sample
in both sample test locations with different clamps, wiring, heaters, and
thermometry and the best fit conductivity agreed to $<4\%$ across the full
temperature range.     

\citet{locatelli1976} and \citet{olson1993} have previously reported the
low--temperature thermal conductivity of Vespel SP--1 (up to 1~K) and SP--22
(up to 2~K).  For SP--22, our results are in very good agreement with
\citet{olson1993} for temperatures below 1~K.  Above 1~K our results favor an
index closer to 1.6.  \citet{locatelli1976} measured SP--1 from 0.1 to 1~K and
we agree with the shallow index of ~1.1--1.2 across this temperature range.
However, our measured conductivity is of order 25--30\% larger than their
result.  Our measurements extend the the thermal conductivity measurements of both types of Vespel up to 4~K.

The thermal conductivity of unfilled PEEK has been measured previously
\citep{gottardi2001}, but we are unaware of previous measurements of
carbon--filled PEEK or glass--filled PEEK at low temperatures.  Our
measurements of the thermal conductivity of unfilled PEEK disagree
significantly with those in \citet{gottardi2001}.  We measure a factor of 2--3
higher conductivity in the region of overlap.  Our PEEK sample is supplied
from a different vendor than that measured in \citet{gottardi2001} and the
thermal conductivity difference may reflect intrinsic differences -- such as
the degree of crystallinity -- between the samples. It is also possible that
differences in experimental setup contributed to the disagreement.  Our own
measurements showed that it is important to isolate the thermometer blocks
from the heaters and heatsinks due to the finite thermal boundary impedance
between the copper clamps and the sample.  

It is interesting to note the similarity in thermal conductivity of unfilled PEEK and 30\% carbon--filled PEEK.  Unlike Vespel, where the addition of graphite significantly changes the conductivity of the material, the addition of carbon fiber to PEEK seems to decrease its conductivity only marginally and the effective index is very similar to unfilled PEEK (see Figure \ref{fig:kt}).  The addition of glass to PEEK appears to affect its conductivity mostly at temperatures above 1~K.

\citet{reed1997} contains a review of the cryogenic properties of composite supports including measurements of the thermal conductivity of carbon fiber composites down to 4~K.  Our measurement of 24 mW/m~K for carbon fiber composite rod at 4~K agrees with the values of $20-36$ mW/m~K in this review. \citet{radcliffe1982} report measurements down to $T{\sim}1.5$~K. Over the range of overlap ($T\sim 1.5-4$~K) our results are in very good agreement with their samples measured along the axis of the fibers for those samples with high fiber volume.  Note that our sample has a fiber volume of 67\% and we measure its thermal conductivity along the fiber direction.  We are not aware of measurements of the thermal conductivity of carbon fiber composites below 1.5~K.

\citet{roth1976} measured Macor ceramic from 0.06~K up to $\sim1.4$~K. Our data are consistent with these results.
This reference also includes the data of \citet{lawless1975} at  temperatures above 2~K (corrected for a misscaling by a factor of 10) and we agree with these data as well.  The thermal conductivity of the base Torlon 4203 PAI has been measured previously \citep{ventura1999}, but we are unaware of published conductivity measurements for Torlon 4301.  

An upper-- and lower--limit to the conductivity of Poco Graphite AXM--5Q is reported in \citet{woodcraft2003}.  The limits of \citet{woodcraft2003} are consistent with our measurements.  We also observe the leveling--off of the thermal conductivity at the low--temperature end.  It is worth noting that the graphite sample is the only material we tested that has an effective index that rises significantly with increasing temperature rather than falling.  As noted in \citet{woodcraft2003}, this feature makes the material attractive for use as a passive heat switch.  With a room--temperature conductivity of around 90~W/m~K, the ratio of room--temperature to $T=1.4$~K conductivity is $\sim$34,000 for AXM--5Q.

\citet{walker1981} report a measurement of the thermal conductivity of G--10 and G--10CR (a ``cryogenic grade" of G--10) along the plane of the glass fibers at temperatures between 0.1 and 4~K.  Our measurements of the thermal conductivity of FR--4 along the plane of glass fibers are visually in good agreement with these results even though our sample contains the halogen flame retardant.  

\begin{figure}[t] 
  \centering
  \includegraphics[bb=82 85 544 713,width=5in,height=6.8in,keepaspectratio]{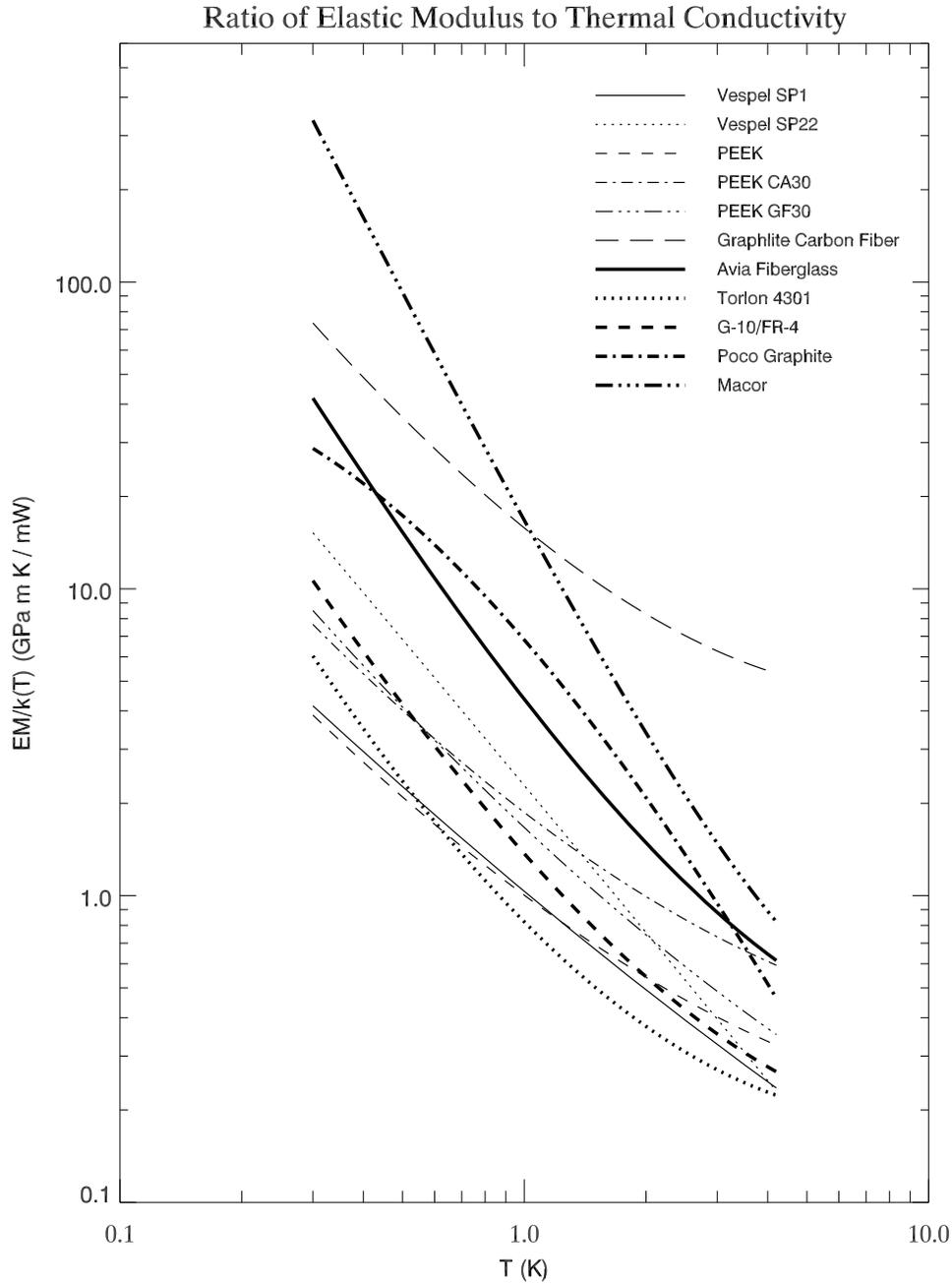}
  \caption{\scriptsize Room--temperature elastic modulus (EM) of the materials divided by the best--fit thermal conductivities for the materials tested.  For oriented fiber composites we have used data along the fiber axis.  We have used the tensile modulus for those materials where a measured compression modulus is unavailable (see Table \ref{matproptable}).  For FR--4, we used the average flexure modulus of 17.5 GPa as a proxy for the elastic modulus and the measured thermal conductivity along the plane of the glass fibers.}
  \label{fig:eoverkt}
\end{figure}

A quality of particular interest to designers of thermally--isolating cryogenic structural supports is the ratio of stiffness to heat flow of a material.  This can be quantified as the ratio of the elastic modulus, EM, to the thermal conductivity, $k(T)$, of a material.  This figure of merit is of particular interest for truss structures where the members are under tension and compression.  It is difficult to come by measurements of the low--temperature moduli of materials.  Although the increase in material stiffness with cooling can be significant for some materials ({\it{eg.}} Teflon), it is informative to plot the room temperature elastic modulus over low--temperature thermal conductivity.  Figure \ref{fig:eoverkt} shows this ratio for the materials tested here.  For fiber-containing composites, we use values of EM and $k(T)$ along the axis of the fibers.

There are a number of noteworthy features in Figure \ref{fig:eoverkt}.  It is worth pointing out to designers on a budget that a number of materials performed better than the two varieties of Vespel tested here.  Two of the materials that performed best (Macor and graphite rod) are brittle and care must be taken to avoid exceeding their rupture strengths.  Note that the flexural strengths of Poco Graphite AXM--5Q (69 MPa) and Macor ceramic (94 MPa) are not far from the ultimate flexural strengths of Vespel SP--1 (110 MPa) and SP--22 (90 MPa).  The carbon fiber rod surpasses the next closest material by nearly an order of magnitude in stiffness--to--conductivity at 4~K.

\section{Conclusions}

We have measured the thermal conductivity of a number of polymeric and
composite materials between 0.3 and 4~K and fit the results to a parametric
model.  We then compare the ratio of the room--temperature elastic modulus and
the low--temperature thermal conductivity to identify candidate materials for
thermally--isolating structural support members.  By this metric, a number of
the inexpensive materials performed better than much more expensive
advanced engineering polymers across the relevant temperature range.  Of
particular note is the pultruded carbon fiber Graphlite rod, which
distinguishes itself at $T\sim1-4$~K, as well as Macor ceramic at low temperatures. 

\section{Acknowledgments}
This work was supported under the \spider project by NASA Grant No. NNX07AL64G.  The authors wish to thank Justin Lazear for his assistance with thermometer cross--calibration, Mike Zemcov for pointing us to Poco Graphite, and Warren Holmes for useful discussions.


\bibliography{thermalcond}

\end{document}